\documentstyle[aaspp4,flushrt]{article}
\def\ltsima{$\; \buildrel < \over \sim \;$}
\def\lsim{\lower.5ex\hbox{\ltsima}}
\def\gtsima{$\; \buildrel > \over \sim \;$}
\def\gsim{\lower.5ex\hbox{\gtsima}}

\newcommand{\etal}{{et al.~}}
\newcommand{\ie}{i.e.\thinspace}
\newcommand{\eg}{e.g.\thinspace}

\begin{document}
\today
\title{How universal are the density profiles of dark halos?}
\lefthead{Huss \etal}
\righthead{How universal are density profiles}

\author{A.~Huss\altaffilmark{1}, B.~Jain\altaffilmark{1,2}, M.~Steinmetz\altaffilmark{1,3}}
\altaffiltext{1}{Max-Planck-Institut f\"ur Astrophysik, Postfach 1523, 85740 Garching,
Germany}
\altaffiltext{2}{Department of Physics and Astronomy, Johns Hopkins University, 
Baltimore, MD 21218, USA}
\altaffiltext{3}{Steward Observatory, University of Arizona, Tucson, AZ 85721, USA}

\centerline{adh@mpa-garching.mpg.de, bjain@pha.jhu.edu, msteinmetz@as.arizona.edu}

\begin{abstract}

We investigate the formation of virialized halos due to the 
gravitational collapse of collisionless matter using high-resolution 
N-body simulations. A variety of formation scenarios are studied, 
ranging from hierarchical clustering to monolithic radial collapse. 
The goal of these experiments was to study departures from the
universal density profiles recently found to arise in cosmological
settings. However, we found that even for models which exhibit quite a
different formation history, the density and velocity dispersion profiles 
of the virialized halos are strikingly similar. 

Power law density profiles do not result even in models with initial 
power law profiles and no initial substructure or non-radial motions. 
Such initial conditions give rise to a radial
orbit instability which leads to curved velocity dispersion and 
density profiles. The shapes of the density profiles in all our
models are well parameterized by the profiles of halos formed  in a 
generic cosmological setting. 
Our results show that the universality of dark halo 
density profiles does not depend crucially on hierarchical merging
as has been suggested recently in the literature. Rather it arises because
apparently different collapse histories produces a near universal 
angular momentum distribution among the halo particles. We conclude 
that the density and velocity dispersion profiles of virialized halos 
in an expanding universe are robust outcomes of gravitational collapse,  
nearly independent of the initial conditions and the formation history. 

\end{abstract}


\section{Introduction}

The hierarchical collapse of dark matter into virialized halos
is likely to have played a key role in the formation of galaxies 
and clusters of galaxies. Several aspects of this process have 
been studied in recent years, in particular the role of initial
conditions and of ongoing mergers in shaping the final structure of 
the dark matter halos. However, it remains an open question whether violent relaxation 
erases all information about the initial conditions and the formation
history of halos. If it does not, then the density  and velocity 
profiles of dark matter halos may bear imprints of
the initial power spectrum as well as the cosmological 
density parameter $\Omega$  and the cosmological constant $\Lambda$.

Such a dependence between the initial power spectrum and the 
density profiles of
virialized objects was pointed out by Hoffman \& Shaham (1985) and Hoffman
(1988). Their conclusions were based on the secondary
infall model of Gunn \& Gott (1972), and the self-similar solution of Fillmore
\& Goldreich (1984) and Bertschinger (1985).  These analytic
calculations, however, were based on simplifying assumptions (\eg spherical
symmetry) and, therefore, could not consider all aspects of gravitational
collapse. N-body simulations, which begin with generic initial conditions,
provide an attractive alternative. Though the effects of gas dynamics are
neglected, they are likely to represent a realistic description of 
the formation of galaxy clusters and the outer parts of galactic halos. 

In recent years, multi-mass simulation techniques have been developed which
allow one to investigate the formation of individual halos with high
numerical resolution. Navarro, Frenk \& White (1996, 1997) 
(hereafter NFW) have
investigated the structure of dark matter halos which form in a cold dark
matter (CDM) universe. They found that the density profiles of
halos do not follow a power law, but tend to have a slope 
$\alpha = d \ln \varrho/d \ln r$ with $\alpha=-1$ near the 
halo center  and $\alpha = -3$ at
large radii. Over more the four orders of magnitude in mass, the density
profiles follow a universal law, which can be parameterized by
\begin{equation}
\frac{\varrho(r)}{\varrho_b} = \frac{\delta_n}{\frac{r}{a_n} (1 + \frac{r}{a_n})^2}.
\label{nfw}
\end{equation}
The two parameters are the scale radius $a_n$ which defines the scale where
the profile shape changes from slope $\alpha >-2$ to $\alpha <-2$, and the
characteristic overdensity $\delta_n$. They are related because the 
mean overdensity enclosed within the virial radius $r_{vir}$ is $180$. 
Equation~(\ref{nfw}) differs slightly 
in its asymptotic  behavior at large radii from the
profile which was proposed by Hernquist (1990) to describe the mass profiles of
elliptical galaxies.  This profile, which goes like $r^{-4}$ at large
radii, was used by Dubinsky \& Carlberg
(1991) to fit the density distribution of halos which were formed in their
CDM-like simulation. 

Cole \& Lacey (1996), Tormen, Bouchet \& White (1996), Navarro, Frenk
\& White (1997), 
Huss, Jain \& Steinmetz 1997 (hereafter HJS) and Moore \etal (1997), 
among others, have
extended the original results of NFW to other initial spectra, to other 
cosmologies  and to 
higher spatial and mass resolution. Most of the above studies show that
halos which form in a variety of cosmological models are
well described by equation~(\ref{nfw}). The scale radius and the central
overdensity appear to be related to the formation time of the halo (Navarro
\etal 1997). These results suggest that the density profile found by NFW is
quite generic for any scenario in which structures form due to hierarchical
clustering. The power spectrum and cosmological parameters only
enter by determining the typical formation epoch of a halo of a given
mass and thereby the dependence of the characteristic radius $a$ on 
the total mass of the halo. 

All these studies were based on initial particle distributions drawn from a
Gaussian random field. Since cold dark matter scenarios have 
initial perturbations on all scales, the formation of halos generically
proceeds by hierarchical merging of dark matter clumps. In this
paper we instead use a set of artificial initial conditions, 
gradually increasing in complexity, to isolate the physical effects
that determine the final properties of the halo. We control the
degree of substructure by varying the velocity dispersion of 
the particles. The merging history of the halos varies between the 
two extremes of monolithic radial collapse and hierarchical merging.  

\section{Models and simulation parameters}

The simulations used in this paper were performed using a N-body code which uses 
the special purpose hardware GRAPE (Sugimoto \etal 1989). The gravitational
force is computed by direct summation over all particles with a Plummer
force law. The Plummer softening parameter is $0.005 h^{-1}\,$Mpc in comoving
units. The initial overdensity $\Delta(x)$ inside a sphere of comoving 
radius $x$, is given by the power law
\begin{equation}
  \label{ovd}
  \Delta(x) =  \Sigma\, (1 + z_i)^{-1} \, 
\left (\frac{x}{8\,h^{-1}\,{\mathrm Mpc}}
  \right )^{-1}.
\label{rhoinit}
\end{equation}
The normalization $\Sigma$ is set to $1.4$, \ie the halo formation
time is the same as in the CDM model  for which the power spectrum is 
normalized to $\sigma_8=0.6$. The size of the sphere at the initial
redshift $z_i=20$ is $15 h^{-1}$Mpc. 
The density field is set up
by radially displacing particles from an initial glass-like particle
distribution (White 1996). According to the Zel'dovich approximation 
(Zel'dovich
1970), the displacement vector is given by $\epsilon(x) = -x/3\, \Delta(x)$.
The initial particle 
velocities correspond to the growing mode of gravitational instability
(in comoving units this is $u(x) = \dot a/a \, \epsilon(x)$). 
In addition varying amounts of initial velocity dispersion are added to the 
particle velocities. Details of the simulation technique used for the 
CDM model are described in an earlier paper (HJS). 

The key parameters of the five models and the halos formed
in each model are given in tables 1 and 2. 
All halos have in excess of $10^4$ particles inside the virial
radius at $z=0$. The masses range from $0.3-0.5\,
\times 10^{15}\,h^{-1}\,M_\odot$
while the virial radii range from $1.1-1.4\, h^{-1}$ Mpc. 

Of the five models listed in table 1, models I-IV use the power
law initial profile of equation \ref{rhoinit}.
Model I is constructed such that it is forced to undergo purely
radial collapse. This is achieved by setting the tangential components
of the gravitational force to zero, and by giving no initial dispersion
in the particle velocities. Model II has the full three dimensional
force but zero initial velocity dispersion. Models III and IV have 
increasing amounts of velocity dispersion given by $\sigma_{3d}$ in table 1.
These two models are designed to study to what extent the evolution of halos is
affected by substructure which is produced by the initial velocity dispersion.
Model V is the standard cold dark matter model. A detailed
study of cluster formation in this model is presented in HJS. 

\begin{table}
\caption{\label{tb1} Description of the five collapse models. 
All models assume h = 0.5; the CDM model is normalized to
$\sigma_8=0.6$. 
$\sigma_{3d}$ is the dispersion in the initial tangential (comoving) 
velocity in km/s.}
\centering
\begin{tabular}{llll}
\hline
Model &   Description & Force & $\sigma$ \\ \hline
I & spherical collapse & radial force & $\sigma_{3d} = 0$\\
II & spherical collapse & 3d force & $\sigma_{3d} = 0$\\
III & spherical collapse & 3d force & $\sigma_{3d} = 200$\\
IV & spherical collapse & 3d force & $\sigma_{3d} = 2000$\\
CDM & Standard CDM & & 
\end{tabular}
\end{table}

\begin{table}
\caption{\label{tb2} Parameters of the halos at $z = 0$. The mass within
the virial radius $M_{vir}$ is in units of $10^{15}\,h^{-1}\,M_\odot$, 
while $r_{vir}$ and $a_n$ are in $h^{-1}$ Mpc.}
\centering
\begin{tabular}{lccccccc}
\hline
Model & $M_{vir}$ & $r_{vir}$ & $\delta_n$ & $a_n$ & $c$ & $N(< r_{vir})$   \\ \hline
I &  0.30 & 1.13 &  - & - & - & 10600 \\
II &  0.33 & 1.16 & 8300 & 0.22 & 5.4 & 11400 \\
III &  0.32 & 1.15 & 6400 & 0.24 & 4.7 & 11100  \\
IV &  0.29 & 1.12 & 11800 & 0.19 & 6.0 & 10200  \\
CDM &  0.50 & 1.36 & 13400 & 0.22 & 6.3 & 10800  
\end{tabular}
\end{table}

\section{Formation history and halo profiles}

Figure 1 compares the formation of halos for the five considered models. Projected particle
positions are shown in regions large enough to include nearly all  
particles. In the purely radial force model collapse proceeds inside out
and a spherically symmetric halo forms in a smooth fashion. In model II which has
no initial inhomogeneity or velocity disperion 
a bar instability develops at redshifts less than $0.5$. By $z=0$ a distinct bar
structure dominates the halo.  This instability appears to be a
radial orbit instability, found in collapse studies in a non-cosmological
context (Polyachenko 1981; Merritt \& Aguilar 1985; Carpintero \& Muzzio 1995). 
We have confirmed the appearance of this bar instability under a variety of initial
conditions. Our results thus demonstrate that the radial orbit instability
can arise in a cosmological setting as well if the initial conditions are
sufficiently smooth. 

The range of models in figure 1 show increasing amounts of merging
as the collapse proceeds. In models III and IV the clumpiness arises due
to the initial velocity dispersion which leads to the aggregation of
particles into small clumps at high redshift. These clumps in turn 
merge to form a single virialized halo at $z=0$. 
The CDM model also shows hierarchical merging, though the collapse 
is more filamentary and occurs at higher redshift. 

The density profiles of the models are shown in figure 2.  We have plotted 
$r^2\, \rho(r)/\rho_b$, where $\rho_b$ is the background density,
in order to show more clearly the curvature of the profile relative
to an $r^{-2}$ power law. Particles were binned such 
that each bin contains at least $100$ particles. 
At high $z$, the density profile shows
a radial infall pattern in all the models, with an $r^{-2}$
profile developing in the inner regions. In model I, which has 
no non-radial forces, the $r^{-2}$ profile remains stable and 
extends to the virial radius at $z=0$. This slope is consistent with
the predictions of the secondary infall model. However in the other 
three spherical models, by $z=0.58$ the profile has developed
significant curvature and is shallower than $-2$ in the inner
regions and steeper near the virial radius. 
By $z=0$ it is evident that with exception of model I all halos show a 
characteristic curved profile. The density profile of the CDM
halo is well fit by the NFW profile. The profiles of the halos
in models II-IV are very similar and are also adequately fit
by the NFW form as shown in figure 2 (see figure 4 and discussion
below for more details). 

The velocity structure accompanying these density profiles at
different redshifts is shown in figure 3. The radial streaming
velocity, $v_r(r)$, indicated by the dashed-dotted line, clearly shows
the infall pattern at high-redshift.  At low redshift, $z\la1$, the
halos become virialized and $v_r(r)$ is indeed consistent with
zero. The velocity dispersions
(radial and tangential shown by the solid and dotted lines, respectively)
are small at $z\simeq 8$. For the lower redshifts, the 
radial dispersion approaches $\simeq 1000 \,$km/s. The tangential
dispersion is nearly the same as the radial dispersion inside 
the scale radius, presumably due to the efficiency of violent
relaxation in the inner regions, but is considerably smaller at larger radii. 
By construction, model I which is shown in the left-most panels,
does not develop any tangential dispersion. 

We have verified the features described above for other models
not shown in figures 1-3. These include power law profiles for the initial
overdensity with the logarithmic
slope varying between $-0.5$ to $-1.5$ and a halo formed in 
a Hot Dark Matter model (see also HJS). In all cases, except for 
the initial conditions of model I,
we recovered density profiles  with characteristic curvature 
and velocity dispersion profiles similar to the ones shown in figure 3. 
Figure 4 shows the density profiles of four halos: three of them
from power law initial profiles and the fourth from CDM initial conditions.
It is evident that there is some scatter (typically $\simeq 20\%$)
about the NFW-fit.
We find that it is possible to fit the measured profiles somewhat better
with a generalized profile, with the inner slope varying between
$-0.5$ and $-1$. This suggests
that the NFW profile is not a unique best fit profile, at least for
halos not formed in a hierarchical cosmology. It does however provide
an adequate fit given the scatter between the different halos. 
Only the halos of model I produced final density profiles that were well 
approximated by power laws in agreement with the analytic 
predictions of the secondary infall model (Fillmore \& Goldreich 1984; 
Bertschinger 1985). 

\section{Discussion}

We have simulated the collapse of dark matter halos with
varying initial conditions. As shown in figure 1 the 
models are designed to vary the amount of sub-structure
and merging in the collapsing halo. We found that 
even with a smooth initial particle distribution and minimal merging,
the virialized density and velocity dispersion profiles of the halos
are very similar to the halos formed in a cold dark matter cosmology. 
The density profile has logarithmic slope $-2$
at a characteristic radius and a slope shallower in the inner
regions and steeper in the outer regions. It is well fit by the
form proposed by NFW. 

The only way we could get a power law 
density profile was to turn off the non-radial components of 
the gravitational force completely. Since this artificial 
condition  can only be satisfied in a simulation, we 
conclude that dark matter halos resulting from gravitational
collapse do not have power law profiles. Instead the density and 
velocity dispersion profiles have a nearly universal characteristic 
shape, independent of the initial conditions or the formation history. 

Our results suggest that the key to universal density profiles
does not lie in the hierarchical merging history of cold dark
matter halos, as recently argued by Syer \& White (1996). 
It is a more generic feature of gravitational collapse, which can be derived
from the near universal behavior of the velocity field of the
halo particles. The velocity dispersion profile grows with time 
in nearly the same way for halos in all the models. 
The mechanism by which angular momentum is
acquired can vary, e.g. model II has little merging and instead shows 
a strong bar instability, but the resulting velocity dispersion 
profiles have similar shape. 

NFW used hierarchical merging as embodied in the
Press-Schechter scenario to model the variation of the scale
radius $a_n$ with halo mass and the cosmological model. To the
extent that their results are empirically verified, they provide
a useful guide to the variations of halo profiles. But our
results suggest that the merging history is not a fundamental
determinant of halo profiles, since the
emergence of the characteristic scale radius occurs even in 
models with negligible merging. It would be interesting to
pursue this issue further with detailed dynamical studies.

\acknowledgments
 
We  wish to thank Julio Navarro, Dave Syer, Bepi Tormen, Scott Tremaine 
and Simon White for stimulating discussions. We are especially grateful to
Bepi Tormen and Simon White for helpful comments on the manuscript.  
This work was partially supported by the
Sonderforschungsbereich SFB 375-95 of the Deutsche Forschungsgemeinschaft.

\begin{figure*}
\epsscale{0.9}
\plotone{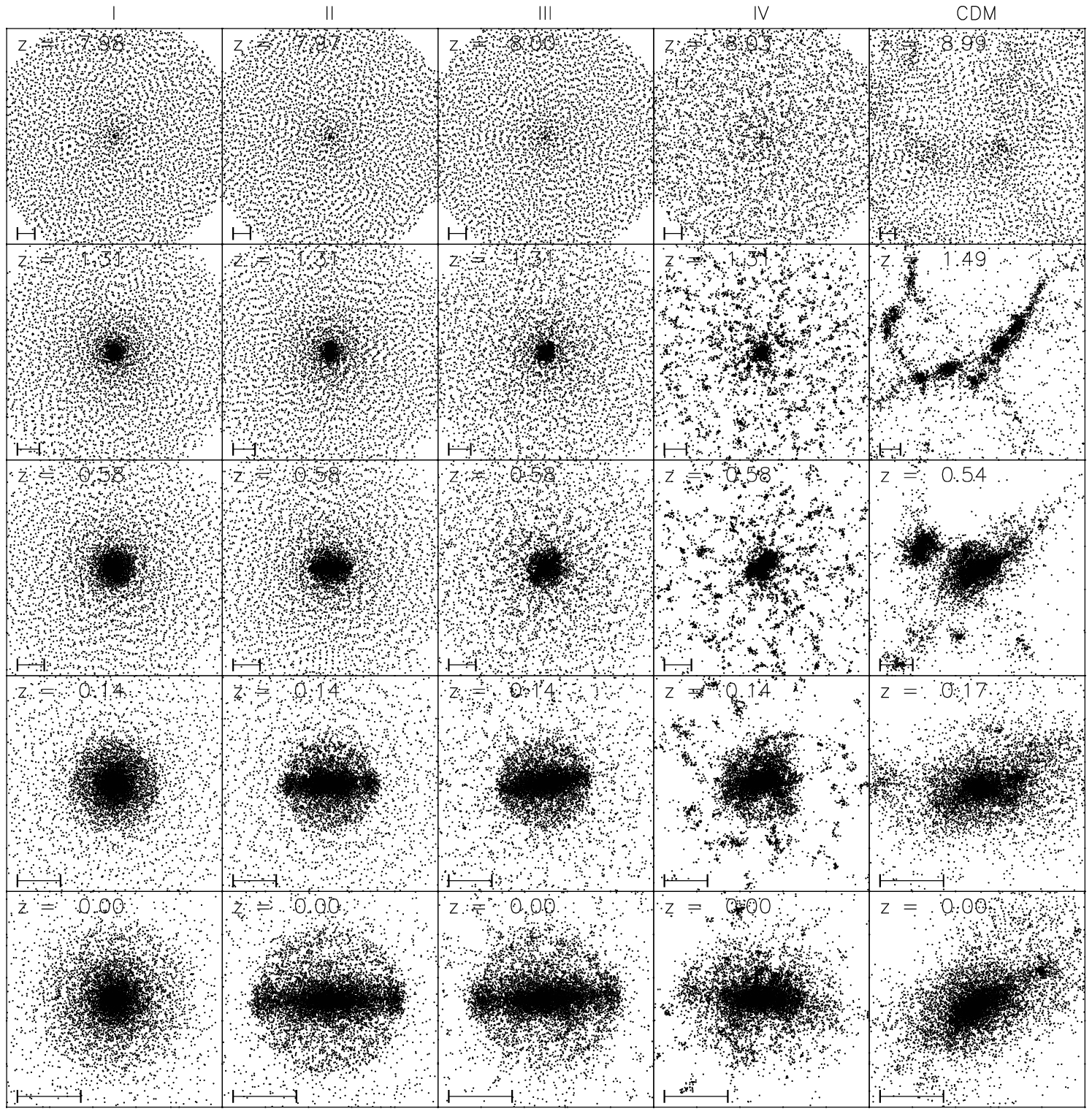}
\vspace{0.5cm}
\caption{\label{figps1}The projected particle distribution of halos at different redshifts 
is shown from top to bottom. The five models (see table 1) are shown from 
left to right. The side length in comoving units for the spherical collapse 
models I-IV is (from the top) $6.2$, $4.9$, $4.0$, $2.5$ and
$1.4\,h^{-1}$ Mpc. For the CDM simulation the side length is
$7.2$, $5.3$, $3.3$, $1.7$ and
$1.4\,h^{-1}$ Mpc. A line of length $1\,h^{-1}$ Mpc is shown
within each box.}

\end{figure*}

\begin{figure*}
\epsscale{0.8}
\plotone{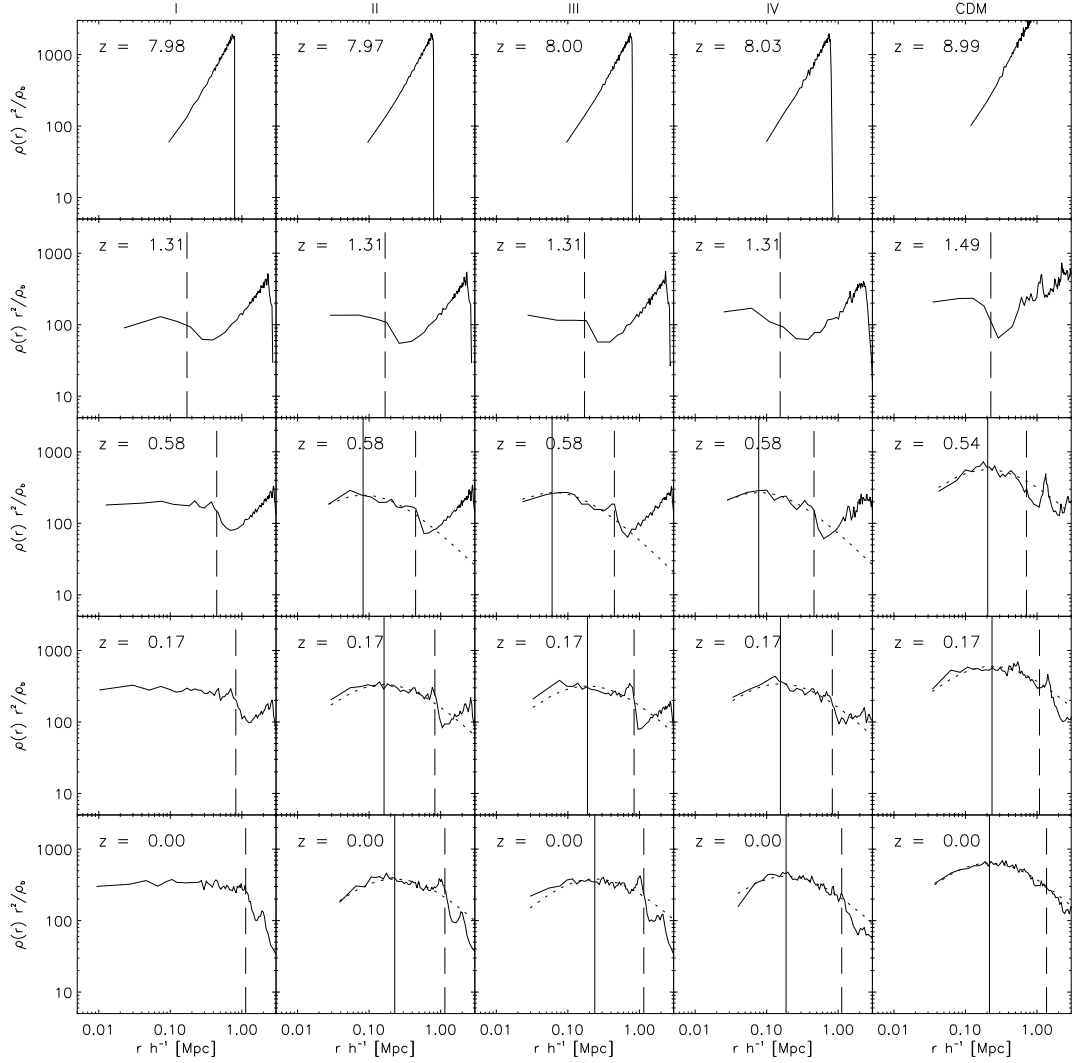}
\vspace{0.5cm}
\caption{\label{figps2}Density profiles shown at the same redshifts and 
for the same models as the particle distribution plot 
in figure~\ref{figps1}. The vertical long dashed line is the virial radius
$r_{vir}$, while the vertical solid line is the scale length $a_n$ of the
NFW profile. The dotted curves show the NFW-profile given in equation 1. 
It is evident that at low redshifts, $z\lsim 1$, the density profiles 
of all the halos except model I are adequately fit by the NFW-profile. }

\end{figure*}

\begin{figure*}
\epsscale{0.8}
\plotone{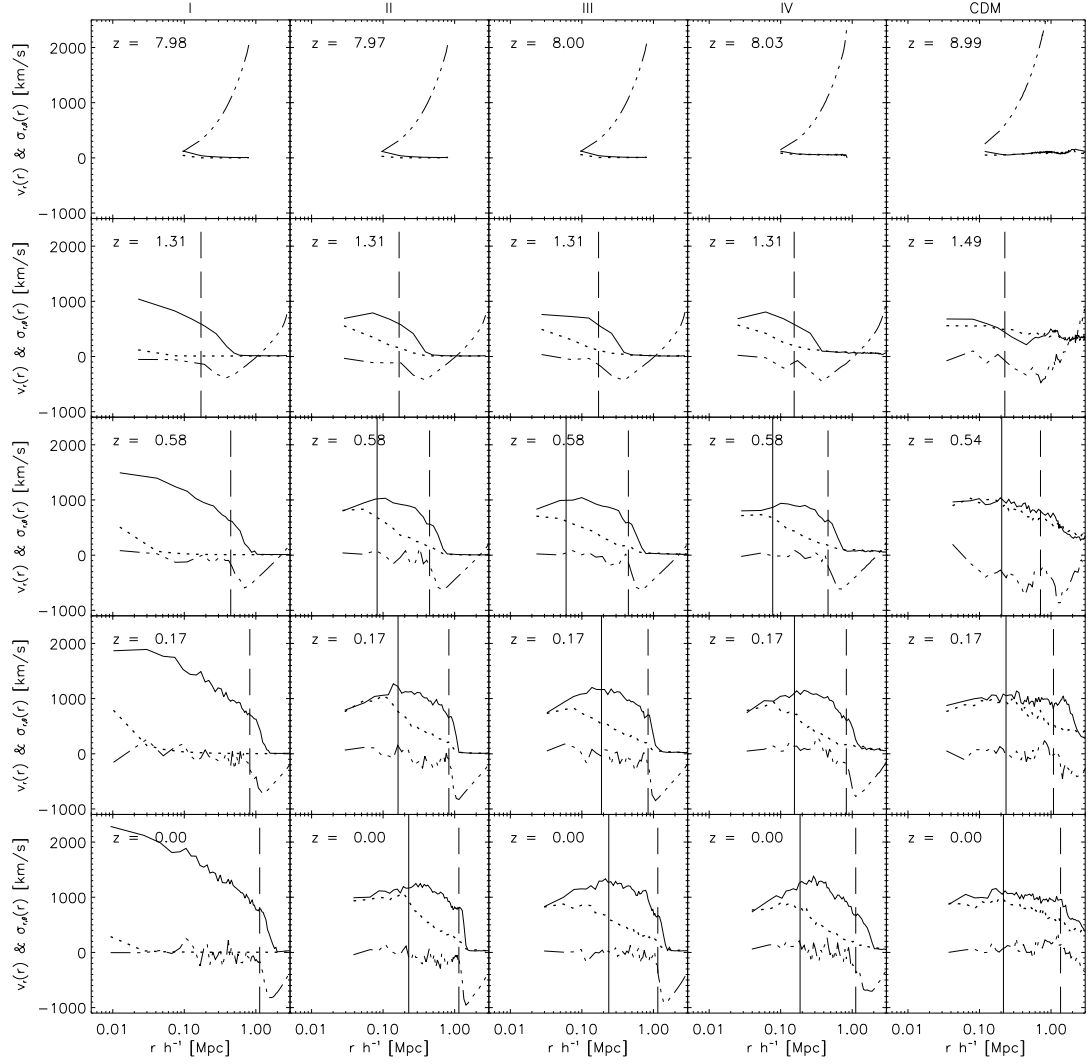}
\vspace{0.5cm}
\caption{\label{figps3}Radial streaming velocity $v_r(r)$ (dashed
triple-dotted line) and the radial (solid line) 
and tangential (dotted line) velocity dispersions, $\sigma_r(r)$ and
$\sigma_t(r)$ as a function of radius $r$. The models and redshifts are the 
same as in the previous
two figures. The vertical long dashed line is the virial radius
$r_{vir}$, while the vertical solid line is the scale length $a_n$
of the NFW profile.}

\end{figure*}

\begin{figure*}
\epsscale{0.8}
\plotone{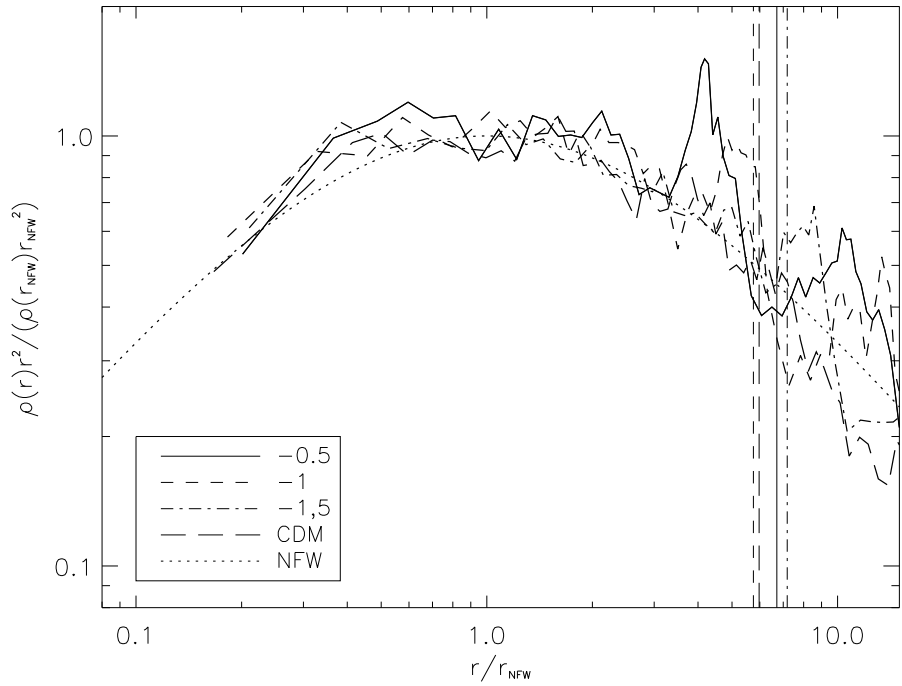}
\vspace{0.5cm}
\caption{\label{figprofiles}Density profiles at $z\simeq 0$ for three different
power law models and the CDM model. The power law models plotted have
initial profiles with logarithmic slopes $=-0.5, -1, -1.5$. The dotted
curve shows the NFW fit, 
while the vertical dashed lines gives the virial radius for each halo.
}

\end{figure*}

\end{document}